# Electrically Tuning Quasi-Bound States in the Continuum with Hybrid Graphene-Silicon Metasurfaces

Ziqiang Cai,[1] Xianzhe Zhang,[1] Tushar Sanjay Karnik,[2] Yihao Xu,[3] Tae Yoon Kim,[1]

Juejun Hu,[2] Yongmin Liu[1,3,*]

[1]Department of Electrical and Computer Engineering, Northeastern University,

Boston, Massachusetts 02115, USA

[2]Department of Materials Science and Engineering, Massachusetts Institute of Technology, Cambridge,

Massachusetts 02139, USA

[3]Department of Mechanical and Industrial Engineering, Northeastern University,

Boston, Massachusetts 02115, USA

[#] Corresponding author. Email: y.liu@northeatern.edu

## Abstract

Metasurfaces have become one of the most prominent research topics in the field of optics owing to their unprecedented properties and novel applications on an ultrathin platform. By combining graphene with metasurfaces, electrical tunable functions can be achieved with fast tuning speed, large modulation depth and broad tuning range. However, the tuning efficiency of hybrid graphene metasurfaces within the short-wavelength infrared (SWIR) spectrum is typically low because of the small resonance wavelength shift in this wavelength range. In this work, through the integration of graphene and silicon metasurfaces that support quasi-bound states in the continuum (quasi-BIC), we experimentally demonstrate significant transmittance tuning even with less than 30 nm resonance wavelength shift thanks to the high quality-factor of quasi-BIC metasurfaces. The tunable transmittance spectrum was measured using Fourier Transform Infrared Spectroscopy (FTIR) with a modified reflective lens to improve the accuracy, and the electrical tuning was realized utilizing the "cut-and-stick" method of ion gel. At the wavelength of 3.0 μm, the measured change of transmittance $\Delta T = T_{\max} - T_{\min}$ and modulation depth $\Delta T/T_{\max}$ can reach 22.2% and 28.9%, respectively, under a small bias voltage ranging from -2 V to +2 V. To the best of our knowledge, this work is the first experimental demonstration of tunable graphene/quasi-BIC metasurfaces, which have potential applications in optical modulation, reconfigurable photonic devices, and optical communications.

## Introduction

Metasurfaces are planar, ultrathin optical components composed of arrays of artificial structures called meta-atoms, with unit cell size and thickness much smaller than the wavelength [1-3]. As the two-dimensional version of metamaterials, metasurfaces can control the amplitude, phase, polarization, and angular momentum of light in a prescribed manner. Many intriguing functions and applications have been demonstrated by metasurfaces, such as nonlinear optics [4-6], planar lenses [7-9], holograms [10, 11], ultrathin polarizers [12, 13], etc. In recent years, BIC metasurfaces have attracted intensive interest [14, 15]. The BIC state represents a non-radiating localized resonant mode that coexist with a continuous spectrum of radiating waves [16]. Since the BIC mode is decoupled from the radiating waves, it has an infinitely large quality-factor (Q-factor). By breaking the symmetry of metasurfaces, the BIC will turn into quasi-BIC with a finite yet high Q-factor, which is inversely proportional to the square of asymmetry factor $\alpha$ [17]. For this reason, we can readily control the Q-factor by modifying $\alpha$, which is especially useful in the studies of strong light-matter couplings [18-21], nonlinear optics [22, 23], molecule detection [24], wavefront shaping [25], etc. People are also interested in tunable quasi-BIC metasurfaces [26, 27], but only few experimental demonstrations have been reported by utilizing phase change materials and electro-optically active polymers [28, 29].

On the other hand, since the first demonstration of isolated high-quality mono-layer graphene through mechanical exfoliation in 2004 [30], the tunable optical properties of graphene have been extensively studied [31, 32]. By incorporating graphene into metasurfaces, people have developed novel devices that can be electrically tuned. These devices offer the advantages of fast tuning speed, high modulation efficiency, broadband tunable electro-optical properties, compatibility with silicon fabrication process, as well as compactness [33-42]. The tunable graphene metasurfaces show great potential in numerous applications, including polarization tuning [43, 44], phase tuning [45-47], photodetectors [48-55], chemical sensing [56, 57], tunable lenses [58-61], etc. However, in most cases, the tuning wavelengths are limited to the mid-infrared or far-infrared range. In contrast, tunable graphene metasurfaces in the SWIR range have been rarely reported, primarily due to the weak intraband transition that determines the resonance wavelength shift in this spectrum [42]. Although it has been reported that critical coupling is achievable by integrating graphene with BIC metasurfaces [62, 63], in-situ tuning of graphene/quasi-BIC metasurfaces has not been demonstrated in experiments.

Here, we report hybrid graphene-silicon metasurfaces to electrically tune quasi-BIC. We first adjust the asymmetry factor α to approach the critical coupling, which is manifested as an absorption close to 50% according to the coupled-mode theory [64]. Measurements are performed with a modified FTIR reflective lens that can suppress beam divergence to achieve the high Q-factor. Next, by utilizing ion gel to provide a large capacitance without diminishing the quasi-BIC resonance, we demonstrate that the transmittance spectra of the quasi-BIC metasurface can be electrically modulated with significant tunability, thanks to the high Q-factor of quasi-BIC. The experiment results agree well with the simulation results.

**Results and Discussions**

As shown in **Figure 1a**, the unit cell of our hybrid graphene-silicon metasurface contains two silicon rectangular nanobars that form a dimer structure, with graphene placed on the top. The silicon metasurface was fabricated through dry-etching of 300 nm silicon layer with a Cr etch-mask. The details about fabrication can be found in Section 1 of the Supplementary Materials. The scanning electron microscope (SEM) image of the fabricated silicon metasurface is presented in the left panel of **Figure 1b**. The difference between lengths of the two nanobars determines the asymmetry factor $\alpha = (L_1 - L_2)/L_1$. At the BIC resonance, two electric dipoles with opposite directions can be excited, denoted as $p_1$ and $p_2$ in the right panel of **Figure 1b**. When $\alpha$ equals 0, the net electrical dipole moment is 0, corresponding to a BIC with an infinite Q-factor. When $\alpha$ is larger than 0, there is a net electrical dipole moment that can couple to the free-space light with polarization along the $y$-axis, corresponding to a quasi-BIC with a finite Q-factor. In the following, the polarization of incident light is always along the $y$-axis if not specified otherwise. As mentioned in the introduction, $Q \propto \alpha^{-2}$. This scaling law provides a simple way to manipulate the Q-factor by modifying the geometry. **Figure 1c** and **1d** show the normalized electric field distributions of quasi-BIC resonance mode at the resonance wavelength of 2.75 μm when $\alpha = 0.2$. The pronounced light confinement around the silicon nanostructures makes the quasi-BIC metasurface an ideal platform to achieve strong graphene-light interactions.

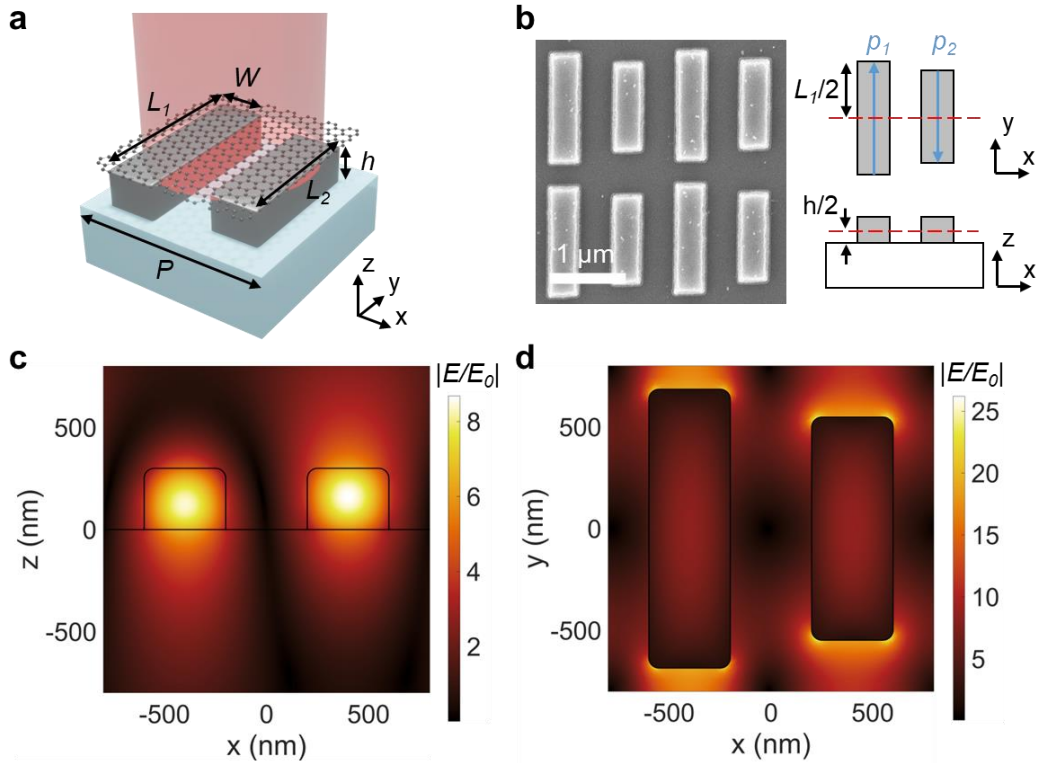

**Figure 1**. Design of the hybrid graphene-silicon metasurface. (a) Schematic of the unit cell. Graphene is placed on top of an array of silicon nanobars, which are fabricated on a silica substrate. The key geometric parameters are: period $P = 1600\ nm$, width of the rectangular bar $W = 400\ nm$ and height $h = 300\ nm$, length $L_1 = 1370\ nm$, and $L_2 = (1 - \alpha) \cdot L_1$, where $\alpha$ is defined as the asymmetry factor. (b) SEM image of the silicon quasi-BIC metasurface without graphene (left), as well as cross-section views of the unit cell (right). The blue arrows in opposite directions in the $xy$-plane are the excited electric dipoles ($p_1$ and $p_2$), and the red dashed lines in the top right and bottom right figures correspond to the positions for electrical field plots in figure (c) and (d), respectively. (c, d) Normalized electric field distributions $|E/E_0|$, where $E_0$ is the input electric field, of the silicon metasurface without graphene at quasi-BIC resonance in the $xz$- and $xy$-planes, respectively. In the two plots, $\alpha = 0.2$ and the wavelength is 2.75

μm.

Since the optical response of quasi-BIC metasurface largely depends on the angle of incidence [65-67], it is crucial to minimize the beam divergence of the incident light. One simple way to reduce the beam divergence is to apply black tape on the reflective lens used in FTIR [68]. **Figure 2a** shows that without the tape, collimated light is reflected by two mirrors and then focused on the sample with a range of incident angles. The measured spectrum is an integrated average for all incident angles. Therefore, the high-Q quasi-BIC resonances cannot be accurately characterized with the original reflective lens, as can be seen in **Figure 2c**. By applying black tape to block light components with large incident angles, the incident light is confined in a narrower range, as illustrated in **Figure 2b**. And the high-Q quasi-BIC resonances are clearly observable from the measured spectra plotted in **Figure 2d**. Noteworthy, we deliberately orient the rectangular window along the $y$-axis, so that the incident light mostly lies in the $yz$-plane. Such an arrangement can help reduce the influence of beam divergence. More discussions on how the incident angle influences the optical response can be found in Section 2 of the Supplementary Materials.

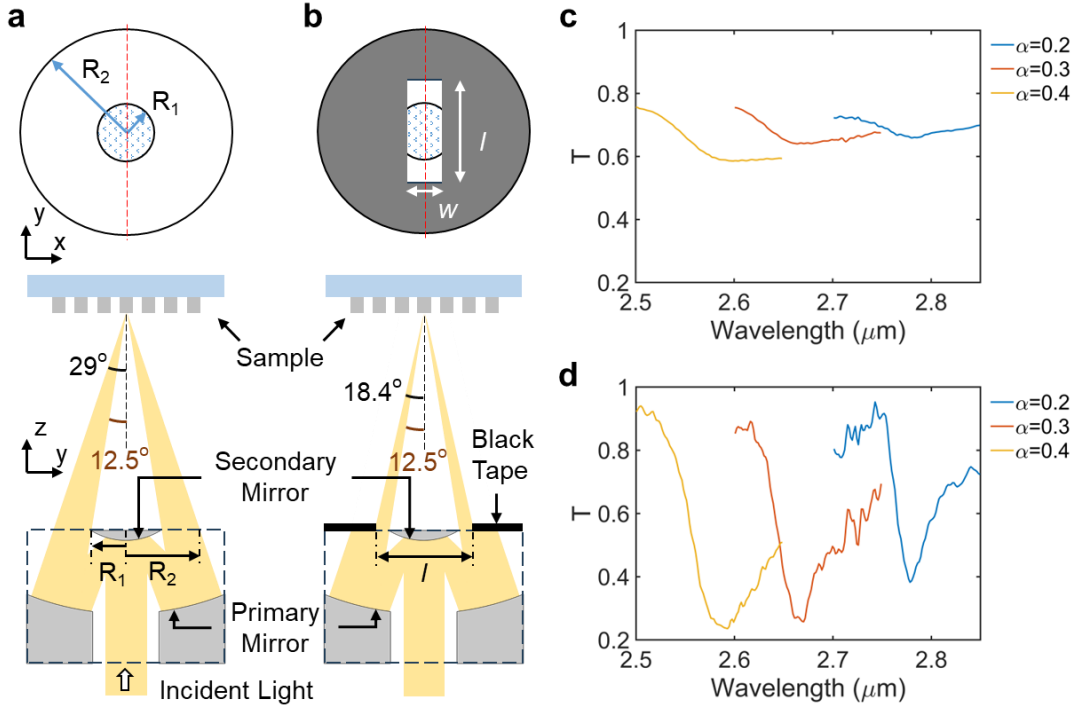

**Figure 2**. FTIR measurement using the original and modified reflective lenses. (a) Schematics of the original reflective lens used in FTIR in the $xy$-plane (top) and $yz$-plane (bottom). Without any modifications, the divergence of the incident beam is $16.5°$. $R_1 = 5.0$ mm and $R_2 = 12.5$ mm, corresponding to the inner and outer radii of the light cone right after the reflective lens. (b) Schematics of the modified reflective lens. After applying a black tape, the portion of the light beam with large incident angles is blocked, and the beam divergence reduces to less than $6°$. The width and length of rectangular window created by the black tape are $w$=5 mm and $l$=15 mm, respectively. The red dashed lines in the top panels of (a) and (b) mark the positions of the cross-section views shown in the bottom panels. (c, d)

Measured transmittance spectra for quasi-BIC metasurfaces with different asymmetry factors $\alpha =$ 0.2, 0.3 and 0.4, using the original reflective lens (c) and the modified reflective lens (d).

After we fabricate the quasi-BIC metasurfaces with asymmetry factors $\alpha = 0.2$, 0.3 and 0.4, we then use FTIR to measure the transmittance ($T$) and reflectance ($R$) spectra. The absorptance ($A$) spectra can be calculated by $A = 1 - T - R$. Based on the coupled mode theory (CMT), it has been shown that the absorptance of a resonant mode in a two-port system can be expressed as [64]:

$$A = \frac{2\gamma_0 \gamma}{(\omega-\omega_0)^2+(\gamma_0+\gamma)^2} \tag{1}$$

Here $\gamma_0$ and $\gamma$ are the intrinsic loss rate and radiative loss rate, respectively. $\omega_0$ is the resonant frequency, and $\omega$ is the frequency of light. The radiative loss $\gamma$ originates from the optical radiation of the resonant mode. It is related to the asymmetry factor following the relationship $\gamma \propto \alpha^2$ in quasi-BIC metasurfaces [17]. The intrinsic loss rate $\gamma_0$, on the other hand, arises from multiple sources, such as the scattering of surface contaminations or edge roughness, ohmic loss from plasmonic resonance, absorption in materials, etc. Under the critical coupling condition, that is, $\gamma_0 = \gamma$, absorptance $A$ reaches its maximum value 50% when $\omega = \omega_0$. For the graphene/quasi-BIC metasurface, $\gamma$ and $\gamma_0$ can be tuned by varying $\alpha$ and the number of graphene layers, respectively. Therefore, the optimal condition for strong coupling and 50% absorptance can be achieved [62, 63].

**Figure 3** compares the simulated spectra with the measurement results, showing good agreement with each other. The simulated reflectance and transmittance spectra for $\alpha = 0.2$, 0.3 and 0.4 are depicted in **Figure 3a**, corresponding to $Q = 108.7$, 51.6 and 27.4, respectively. The measured reflectance and transmittance spectra are shown in **Figure 3b** with Q-factors equal to 90.5, 53.0 and 32.5, respectively, which are extracted from Fano fitting. The measured Q-factor for $\alpha = 0.2$ is smaller than the simulation result, which is likely due to the lower accuracy of FTIR for measuring high-Q BIC resonance. **Figure 3c** and **3d** plot the simulated and measured absorptance spectra after transferring different layers of graphene onto the quasi-BIC metasurface. As discussed before, the critical coupling is reached when $\gamma = \gamma_0$, resulting in the maximum absorptance of 50%. When $\alpha$ increases from 0.2 to 0.4, radiative loss rate $\gamma$ increases, requiring more layers of graphene to approach the critical coupling. This trend has been confirmed in both simulation and experiment. As presented in **Figure 3c**, with monolayer graphene, the peak absorptance decreases when $\alpha$ increases, indicating a weaker coupling between graphene and the quasi-BIC resonance. After increasing the layers of graphene, the absorptance for $\alpha = 0.3$ and 0.4 increases but is still lower than $\alpha = 0.2$ case, as illustrated by the dashed lines in **Figure 3c**. Measured absorptance spectra are shown in **Figure 3d**, in which $\alpha = 0.2$ with monolayer graphene shows the highest absorptance, agreeing with the simulation results. In the experiment, more layers of graphene are needed to reach the absorptance level comparable to simulation results. Specifically, for $\alpha = 0.3$, it takes 2 layers of graphene to achieve about 30% absorptance in simulation, while in experiment, it requires 3 layers. For $\alpha = 0.4$, the simulation shows that the absorptance can reach more than 25% with 3 layers of graphene, whereas the experiment requires 5 layers to achieve comparable absorptance. One possible reason is that during the graphene transfer process, there are always polymer residues left on top of graphene, which increases the distance between the subsequently transferred graphene and the metasurface, leading to reduced coupling between graphene and quasi-BIC resonance. Such a phenomenon has also been observed in our previous work [42], and is more significant with a large number of graphene layers, since more polymer residues are accumulated on the sample surface. In conclusion, the quasi-BIC metasurface with $\alpha = 0.2$ has the strongest interaction with monolayer graphene, which is verified by both simulation and experiment results.

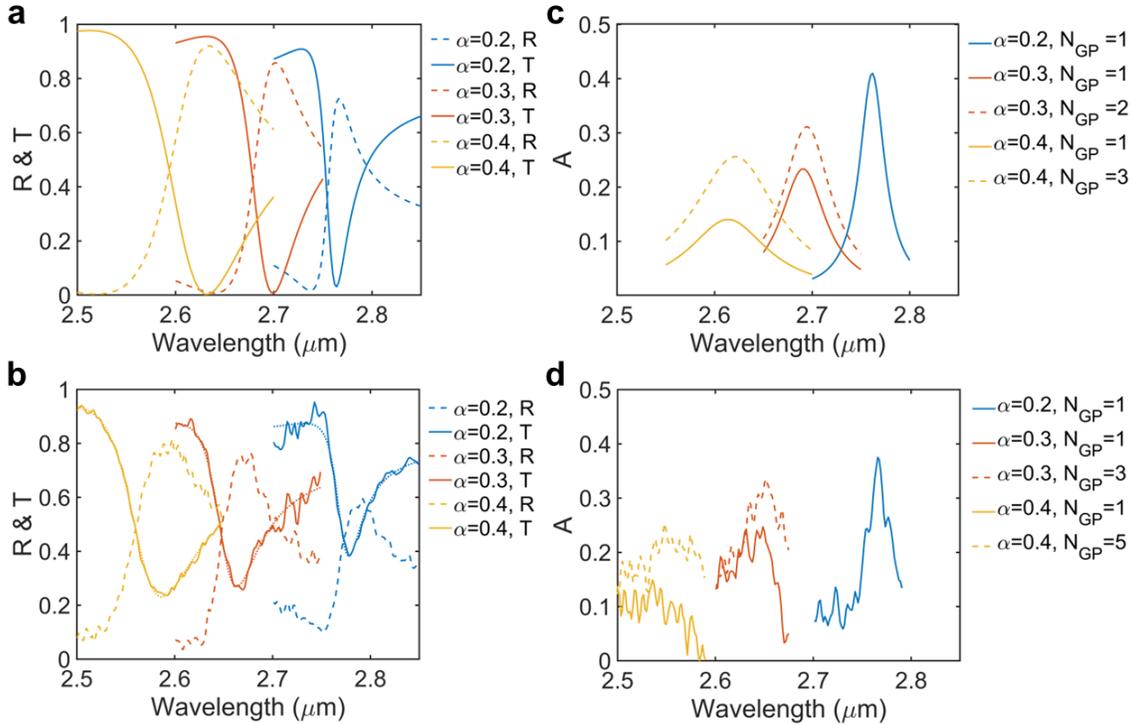

**Figure 3**. Reflectance and transmittance spectra of the quasi-BIC metasurface, as well as the absorptance spectra after graphene transfer. (a) Simulated reflectance and transmittance spectra of the quasi-BIC metasurface. The Q-factor decreases when $\alpha$ increases. (b) Measured reflectance and transmittance spectra of the quasi-BIC metasurface. The dotted lines are the fitting curves for transmittance using the Fano resonance formula. (c) Simulated absorptance spectra with graphene added on top of the metasurface. $N_{GP}$ denotes the number of graphene layers. (d) Measured absorptance spectra after transferring graphene onto the metasurface.

To study the electrical tuning performance of the hybrid graphene-silicon metasurface, we applied ion gel, a transparent polymer that can provide extremely large capacitance [69-74]. It can be easily fabricated by spin-coating and implemented on various electronic devices, including electrically tunable graphene metasurfaces [75-77], using the "cut-and-stick" method. Characterization details of the electrical and optical properties of our synthesized ion gel can be found in Section 3 of the Supplementary Materials. **Figure 4a** illustrates how we use the ion gel to electrically tune the hybrid metasurface, and **Figure 4b** shows the fabricated hybrid metasurface with electrodes. By applying a bias voltage $V_g$, ions with opposite charges accumulate around the source and gate electrodes, forming a nanometer-thick electric double layer that functions as a capacitor [74]. **Figure 4c** demonstrates measured transmittance spectra when $V_g$ is ±2 V. During the measurement, we decrease the sizes of the rectangular window defined by black tape, which can further minimize the divergence of incident light while sacrificing the signal to noise ratio. Specifically, the dimensions of the rectangular window are decreased to $w$=2.5 mm and $l$=12 mm, which reduces the beam divergence to 2.4°. More discussions on how the window sizes influence the measurement results can be found in Section 2 of the Supplementary Materials. Extracted from the Fano fitting curves, the measured transmittance decreases from 76.8 % to 54.6 % when $V_g$ changes from +2 V to -2 V at 3.0 μm. Corresponding $\Delta T$ and $\Delta T/T_{max}$ equal to 22.2% and 28.9%, respectively. **Figure 4d** presents the simulation results. As shown in this figure, when the Fermi energy $E_F$ increases, the resonance wavelength ($\lambda_{res}$) decreases due to a change of graphene's conductivity, which can also be observed in **Figure 4c** considering a negative $V_g$ can increase the doping level of graphene. This 'blue-shift' of $\lambda_{res}$ contributes to the spectral tuning of our hybrid graphene-silicon metasurface. We have also noticed that the measured transmittance tuning is smaller than the simulated one, which is due to the finite beam divergence in our experiments. Other factors that can contribute to the difference between experiment and simulation are fabrication defects, finite thickness of ion gel, light scattering by the edge of black tape, etc.

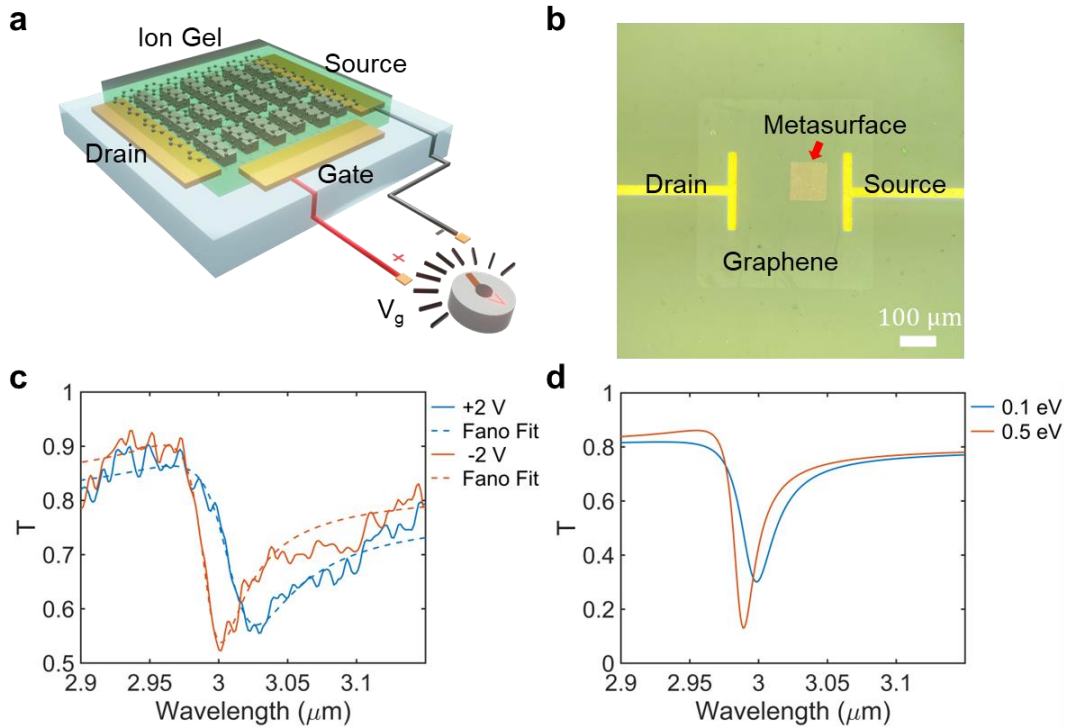

**Figure 4**. Electrical tuning of the hybrid graphene-silicon metasurface. (a) Schematic of the electrical tuning setup. A layer of ion gel is placed on the top. There are three Au electrodes, which serve as source, drain and gate, respectively. The source and gate electrodes are connected to a voltage source to provide a bias voltage $V_g$. (b) Optical microscope image of the metasurface with electrodes. The gate electrode is out of view in this image. (c) Measured transmittance spectra under different $V_g$. The dashed lines are the fitting

curves based on the Fano resonance formula. (d) Simulated transmittance tuning when $E_F = 0.1$ and 0.5 eV. In the simulation, the incident angle is 15° without beam divergence.

To further explore the tuning behavior of our device, we measured the transmittance spectra by gradually changing $V_g$ from +2 V to -2 V with 1 V in each step, as shown in **Figure 5a**. The resonance wavelength, at which the transmittance shows the minimum, and the Q-factor for each bias voltage can be extracted from the Fano fitting curves, as presented in **Figure 5b** and **5c**, respectively. The measured resonance wavelength shift ($\Delta\lambda_{res}$) is 25.8 nm when $V_g$ is tuned from -2 V to +2 V. Such a small $\Delta\lambda_{res}$ for graphene metasurfaces in SWIR has also been observed in our previous work [42]. Although it is much smaller than the $\Delta\lambda_{res}$ of graphene metasurfaces designed in mid-infrared range, which is typically several hundreds of nanometers [47, 78], tuning of transmittance is still substantial because of the enhanced Q-factor thanks to the quasi-BIC resonance. The Q-factor also changes with the bias voltage, as inferred from **Figure 5c**. When $V_g$ is negative, the corresponding $E_F$ is high, which results in lower light absorption of graphene due to Pauli exclusion principle. The Q-factor is mainly determined by the defects in silicon nanostructures in this range and is weakly related to $V_g$. When $V_g$ increases to a positive value, the corresponding $E_F$ becomes low, leading to higher light absorption of graphene due to the interband transition. The Q-factor is primarily determined by absorption of graphene in this case, which can be readily controlled by $V_g$.

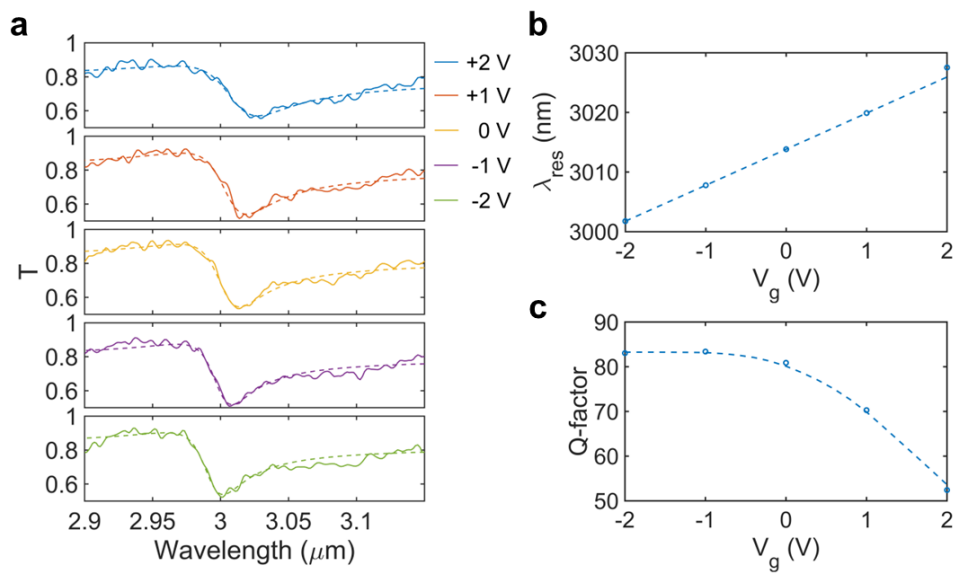

**Figure 5**. Details of the tuning performance. (a) Measured transmittance spectra at different $V_g$. The dashed lines are the fitting curves based on the Fano resonance formula. (b) Resonance wavelengths at different $V_g$. (c) Q-factors at different $V_g$. The Dashed lines in (b) and (c) are guide for eye.

We have also measured the dynamic response of our electrically tunable hybrid graphene-silicon metasurface using a quantum cascade laser (QCL). The response time is around 3 seconds, which is slow due to the long polarization relaxation process in ion gel [79]. Note that the intrinsic response time of graphene can be down to several picoseconds [80]. More details about the time-domain measurements can be found in Section 4 of the Supplementary Materials.

**Conclusion**

In summary, we have experimentally demonstrated, for the first time, an electrically tunable graphene/quasi-BIC metasurface in the SWIR range. When the asymmetry factor $\alpha$ was varied from 0.2 to 0.4, the quasi-BIC metasurface with $\alpha = 0.2$ exhibited the strongest coupling with graphene. Measurements were conducted via a modified FTIR reflective lens to suppress beam divergence, allowing us to accurately characterize the metasurfaces. Afterwards, we tested the electrical tuning utilizing ion gel. Owing to the high Q-factor of the quasi-BIC resonance, we observed substantial tuning of the optical response, even with less than 30 nm shift of the resonance wavelength. Around the wavelength of 3.0 µm, the measured transmittance change $\Delta T$ and modulation depth $\Delta T/T_{\max}$ were 22.2% and 28.9%, respectively, under very modest bias voltages. All the experiment results agreed with simulations. We believe that our work can pave the way for tunable quasi-BIC metasurfaces and broaden their applications in optical modulation, reconfigurable photonic devices, advanced sensing, and beyond.

**Acknowledgements**

Y. L. acknowledges the financial support from DARPA under award number HR00112320025.

**Methods**

*Numerical Simulation*: Graphene-light interactions typically involve interband transition and intraband transitions [31, 81, 82]. The optical conductivity of graphene can be calculated by the random phase approximation [31, 82]:

$$\sigma(\omega) = \frac{i2e^2 k_B T}{\pi \hbar^2 (\omega + i\tau^{-1})} ln\left[2\, cosh\left(\frac{E_F}{2k_B T}\right)\right] + \frac{e^2}{4\hbar}\left[\frac{1}{2} + \frac{1}{\pi} arctan\left(\frac{\hbar\omega - 2E_F}{2k_B T}\right) - \frac{i}{2\pi} ln\frac{(\hbar\omega + 2E_F)^2}{(\hbar\omega - 2E_F)^2 + 4(k_B T)^2}\right] \quad (2)$$

In this equation, $k_B$ is the Boltzmann constant, $T$ is the temperature, and $\tau^{-1}$ is the damping rate. The first term describes intraband transition, while the second term describes interband transition. Carrier mobility $\mu = 1000 \text{ cm}^2/(\text{V} \cdot \text{s})$ is taken in simulation, which is retrieved from experiment data (see Section 3 in the Supplementary Materials).

Simulations are conducted by commercial software COMSOL Multiphysics, in which graphene is modeled using transition boundary condition. To introduce the contribution of fabrication imperfections to the intrinsic loss rate $\gamma_0$ as material absorption in silicon [83], a small imaginary part is assigned to the refractive index of silicon: $n_{Si}$=3.4302+0.005i, so that the best agreement between simulation and experiment can be obtained, and the real part comes from experiment data [84]. For the electrical tuning simulation, ion gel is considered as a semi-infinite dielectric layer with refractive index $n_{gel}$ equal to 1.45.

*Fano Fitting*: The measured spectra can be fitted to the Fano resonance formula [17]:

$$T = \frac{T_0}{1+q^2} \frac{(q+(\omega-\omega_0)/\gamma)^2}{1+((\omega-\omega_0)/\gamma)^2} + T_{bg} \quad (3)$$

In this equation, $q$ is the Fano asymmetry parameter, $T_0$ and $T_{bg}$ are from the background contribution. $\omega_0$ is the resonance frequency and $\gamma$ is the loss rate. The Q-factor can be calculated by $Q = \omega_0/2\gamma$.

*FTIR Measurement*: Measurements are performed using a Bruker Vertex 70 FTIR coupled with a HYPERION 1000 microscope. For the transmittance measurement, we first measure the transmission through the silicon metasurface with an adjusted aperture to collect only the light passing through the metasurface. Then we move the sample to measure the reference spectrum from the silica substrate. Transmittance is obtained by normalizing the first spectrum with the second spectrum. Reflectance is measured following a similar process but replacing the silica substrate with a gold mirror as the reference.